\documentclass[%
 reprint,
superscriptaddress,
 amsmath,amssymb,
 aps,
]{revtex4-2}

\usepackage{graphicx}
\usepackage{dcolumn}
\usepackage{bm}

\def \be {\begin{equation}}
\def \ee {\end{equation}}

\def\O{{\cal O}}

\def\ep{\epsilon}

\def\d{\partial}

\def \r {{\bf r}}

\def\kp{{k_\perp}}

\begin{document}


\title{The Density of Relic Neutrinos Near the Surface of Earth}

\author{Andrei Gruzinov}
\altaffiliation[Permanently at ]{Center for Cosmology and Particle Physics, Physics Department, NYU.}
\affiliation{Weizmann Institute of Science, Rehovot, Israel}
\author{Mehrdad Mirbabayi}%
\affiliation{International Centre for Theoretical Physics, Trieste, Italy}

\vspace{0.5cm}

\date{\today}

\begin{abstract}

  It has been claimed that matter effects cause an asymmetry in the density of relic neutrinos versus antineutrinos near the surface of Earth, of order $\O(G_F^{1/2})\sim 10^{-4}$,  with the vertical extent $\sim 10$m. We argue that the effect is of order $\O(G_F)\sim 10^{-8}$, with the vertical extent $\sim 1$mm.

\end{abstract}

\maketitle


\section{\label{sec:intro} Introduction}

Direct detection of relic neutrinos (cosmic neutrino background) must be just a matter of time -- one day it will be achieved and the neutrino density near the surface of Earth will be measured. 

An unexpectedly large effect of Earth on the relic Dirac neutrinos was found in \cite{arva}. It has been claimed that relic electron neutrinos and $\mu$ and $\tau$ antineutrinos have an overdensity near the surface of Earth
\be
\frac{\delta 
\rho}{\rho}\sim \frac{\sqrt{m_\nu V}}{T},
\ee
where $m_\nu\sim 0.1$eV is the neutrino mass, $T=1.68\times 10^{-4}$eV is the temperature that appears in the Fermi-Dirac distribution of cosmic neutrinos $f(k)=1/(1+e^{k/T})$, and 
\be
V\sim G_F\frac{\rho _E}{m_p}
\ee
is the neutrino potential inside Earth; $\rho_E$ is the mass density near the surface of Earth. According to \cite{arva} this $O(G_F^{1/2})\sim 10^{-4}$ overdensity has the vertical extent of $\sim 10$m.

We will show that in fact there is an {\it underdensity}, of order $O(G_F)\sim 10^{-8}$, with the vertical extent $\sim 1$mm, if one follows \cite{arva} and treats different neutrino flavors as independent non-relativistic particles.

A more realistic approach would be to work with a $3\times 3$ non-diagonal potential in the neutrino mass basis. But since this does not change the qualitative conclusion, we stick to a simple toy model with only electron neutrinos, within which the precise value of density contrast is derived in \S\ref{sec:Earth}: at the surface of a large sphere, with uniform density $\rho_E$ and constant atomic and mass numbers $Z$,$A$ inside,
\be\label{eq:delfd}
\frac{\delta 
\rho}{\rho}=\mp\frac{\ln 2}{3\zeta(3)}\frac{m_\nu V_0}{T^2},
\ee
where upper (lower) sign refers to neutrinos (anti-neutrinos), and \cite{langa}
\be\label{eq:vex}
V_0=\frac{3\frac{Z}{A}-1}{2\sqrt{2}} G_F\frac{\rho _E}{m_p}
\ee
giving
\begin{eqnarray}
\frac{\delta \rho}{\rho}|_{\rm water}=\mp\frac{\sqrt{2}\ln 2}{18\zeta(3)}\frac{m_\nu}{m_p}\frac{G_F\rho _{\rm water}}{T^2}\\ =\mp 8.5\times 10^{-9}\times \frac{m_\nu}{0.1 {\rm eV}}. \nonumber
\end{eqnarray}

The paper is organized as follows:
\begin{itemize}

    \item In \S\ref{sec:theo} we prove a ``Density Theorem" for Maxwell-Boltzmann particles, showing that for a potential barrier there is an $\O(G_F)$ underdensity. We then prove that for Fermi-Dirac, there is also an {\it underdensity}.

    \item In \S\ref{sec:Earth} we apply the thermal method of \S\ref{sec:theo} to the Earth potential to calculate the neutrino and anti-neutrino density at order $\O(G_F)$, obtaining Eq.\eqref{eq:delfd}.

    \item In \S\ref{sec:pert} we show that the first-order perturbation theory agrees with the thermal method.

    \item In \S\ref{sec:flat} we discuss the flat-Earth approximation. It, too, agrees with the thermal method.

    \item In \S\ref{sec:con} we list the sources of discrepancy between \cite{arva} and our conclusion.
    
\end{itemize}

\section{\label{sec:theo} The ``Density Theorem"} 

In thermal equilibrium, the density of a non-relativistic classical gas in a potential $V(\r)$ is given by $n(\r) = n_0 \exp(-V(\r)/T)$ assuming the potential vanishes at infinity, where the gas density is $n_0$. For a {\em collisionless} gas with Maxwell-Boltzmann distribution at infinity, we still have $n(\r) = n_0 \exp(-V(\r)/T)$, as long as the potential energy is positive and has no local minima. This is because in such potentials all trajectories come from infinity and the equilibrium distribution function depends only on the particle energy \cite{land}.

By the same argument collisionless, non-degenerate quantum particles with Maxwell-Boltzmann distribution at infinity and $V({\bf r})\geq 0$  will be in thermal equilibrium everywhere. However, unlike the classical case the density at a point $\r$ depends not just on $V(\r)$ but also on the behavior of the potential in its vicinity. In an arbitrary potential, one can compute the thermal particle density from the density matrix equation \cite{feyn}:
\be
n({\bf r})=n_0(2\pi \beta)^{3/2}\rho(\beta;{\bf r},{\bf r}),
\ee
\be\label{eq:dif}
\left( \partial_\tau -\frac{1}{2}\nabla_{\bf r}^2+V({\bf r})\right) \rho(\tau; {\bf r},{\bf r}_1)=0,
\ee
\be
\rho(0; {\bf r},{\bf r}_1)=\delta({\bf r}-{\bf r}_1);
\ee
$\beta=1/T$ is the inverse temperature, the particle mass $m=1$. 

Eq.(\ref{eq:dif}) describes diffusion with multiplication. To compute the density $n({\bf r}_0)$ one can do an integral over all paths ${\bf r}(\tau)$ which start and end at ${\bf r}_0$:
\be
{\bf r}(0)={\bf r}(\beta)={\bf r}_0.
\ee
Going along each path, initially equal weights $w$ are destroyed at a rate $V$:
\be
\frac{dw}{d\tau}=-V({\bf r}(\tau))w.
\ee

The weight $w$ cannot be destroyed faster than at a rate $V_{\rm max}$, the  weight $w$ cannot be destroyed slower that at a rate $V_{\rm min}$, and the diffusion with multiplication interpretation immediately leads to the following ``Density Theorem'' :
\be
e^{-\frac{V_{\rm max}}{T}}<\frac{n({\bf r})}{n_0}<e^{-\frac{V_{\rm min}}{T}},
\ee
a result that is manifest from the path integral representation of thermal density 
\be
n({\bf r}_0) =\int_{\r(0) =\r_0}^{\r(\beta) = \r_0} D\r(\tau) \
e^{-\int_0^{\beta} d\tau \left[\frac{1}{2}|\dot \r|^2+V(\r)\right]},
\ee

If neutrinos were Maxwellian at infinity, then, rigorously 
\be
e^{-\frac{V_{\rm max}}{T}}-1<\frac{n-
n_0}{n_0}<0,
\ee
which is an underdensity not greater than $\O(G_F)$.

The Fermi-Dirac distribution, can be represented as a superposition of Maxwell-Boltzmann distributions of different temperatures:
\be\label{eq:FDMB}
\frac{1}{e^{\frac{k}{T}}+1}=\int\limits_0^\infty d\alpha~F(\alpha)~e^{-\frac{k^2}{2\alpha T^2}},
\ee
where the weight function
\be\label{F}
F(\alpha)=(2\pi \alpha)^{-1/2}\sum\limits_{n=1}^\infty (-1)^{n+1}~n~e^{-\frac{1}{2}n^2\alpha}
\ee
is positive at all $\alpha$. Since there is an underdensity in a potential barrier for a Maxwell-Boltzmann of any temperature, it follows that a potential barrier creates an {\it underdensity} for the Fermi-Dirac distribution too. 

\section{\label{sec:Earth}Density near Earth}
When $|V(\r)|\ll T$, the density equation \eqref{eq:dif} can be solved perturbatively. In a spherical potential $V(\r)= V_0 \theta(r_E-r)$, with $r_E$ the Earth radius, the density contrast on the surface is given by
\be\label{eq:drhoMB}
\frac{\delta\rho}{\rho} = -\frac{V_0}{2 T},
\ee
as long as $r_E \gg \sqrt{1/T}$: on average, the path-integral trajectory spends half the ``time'' inside Earth. More generally, particle density is affected by the potential within a diffusion length $\sqrt{1/T}$.

Neutrinos have a Fermi-Dirac distribution with $T\gg 1/r_E$ and $\sqrt{V_0}$. In their case, Eq.\eqref{eq:FDMB} helps express $\delta\rho$ and $\rho$ as integrals over Maxwell-Boltzmann results with temperature $\alpha T^2$. The integral is dominated by $\alpha \sim 1$. Therefore, we can substitute $\rho_\alpha\propto \alpha^{3/2}$ and $\delta\rho_\alpha$ from \eqref{eq:drhoMB}, to find
\be\label{eq:drhoFD}\begin{split}
\frac{\delta\rho}{\rho} =- \frac{V_0}{2 T^2}
\frac{\int_0^\infty d\alpha~\alpha^{1/2}F(\alpha)}
{\int_0^\infty d\alpha~\alpha^{3/2}F(\alpha)}
= -\frac{\ln 2}{3\zeta(3)}\frac{V_0}{T^2},\end{split}
\ee
with veritical extent $\sim 1/T \sim 1$mm. 

So far, the discussion of collisionless particles was restricted to those with non-negative potential because it guarantees the absence of bound states. In the collisionless limit, the distribution of bound states is not fixed by the condition at infinity, but it depends on the formation history of the potential well.

In addition, the expansion \eqref{eq:FDMB} includes arbitrarily low Maxwell-Boltzmann temperatures, $\alpha T^2$. When $\alpha T^2 < V_0$, our thermal method, which is valid for a {\em non-degenerate} quantum gas, is inapplicable because it predicts exponentially large occupation for the bound states. So in full generality, we cannot prove a density theorem for antineutrinos. 

Nevertheless, we claim the result for antineutrino over-density is the same as \eqref{eq:drhoFD} up to a sign. First, because the density of antineutrino bound states in Earth is $\O(V_0^{3/2})$, and any assumption about their occupation number (which is $\leq 1$) has no effect at $\O(V_0)$. Hence, for Maxwellian distribution of unbound states, we can safely apply the thermal method if $T>V_0$ and get \eqref{eq:drhoMB} up to a sign.

In the Fermi-Dirac case, exponentially large occupations would be avoided if we cut off the integral \eqref{eq:FDMB} at some $\alpha_{\rm min}\sim V_0/T^2$. This would introduce an error of $\O(V_0^2)$ in $\rho$, which is again subdominant. 

Below, we will see that perturbation theory and flat-Earth approximation both reproduce the same result. 
\section{\label{sec:pert}Perturbation Theory}
For small enough $V_0$, one can compute $\delta\rho$ by a straightforward application of first-order perturbation theory. To be absolutely sure, we have also confirmed it by a brute-force numerical simulation, for a sphere, with and without bound states. 

We first compute energy eigenstates, $E=\frac{1}{2}k^2$, in a potential $V({\bf r})$,
\be
-\frac{1}{2}\nabla^2\psi+V\psi=\frac{k^2}{2}\psi .
\ee
The unperturbed state is
\be
\psi _0=e^{i{\bf k}\cdot{\bf r}},
\ee
and the perturbation, to first order, satisfies
\be
(\nabla^2+k^2)\delta \psi=2V\psi_0,
\ee
giving
\be
\delta \psi({\bf r})=-\frac{1}{2\pi}\int d^3r'~\frac{e^{ikR}}{R}~V({\bf r}')\psi_0({\bf r}'),
\ee
\be
{\bf R}\equiv {\bf r}-{\bf r}'.
\ee

The density perturbation, to first order,
\be\begin{split}
\delta |\psi(\r)|^2 & =\psi_0^*(\r)\delta\psi(\r)+\psi_0(\r)\delta\psi^*(\r)\\[10pt]
&=-\frac{1}{2\pi}\int d^3R~\frac{e^{ikR-i{\bf k}\cdot{\bf R}}}{R}~V({\bf r}-{\bf R})+{\rm c.c.}\end{split}
\ee

For an arbitrary isotropic distribution over momenta $f(k)$, the overdensity is
\be
\frac{\delta \rho ({\bf r})}{\rho}=-\int d^3R~G(R)~V({\bf r}-{\bf R}),
\ee
\be
G(R)=\frac{1}{2\pi R^2}\frac{\int dk ~kf(k)~\sin(2kR)}{\int dk ~k^2f(k)}.
\ee
  
For Maxwell-Boltzmann $f(k)\propto e^{-\frac{k^2}{2T}}$,
\be
G_{\rm MB}(R)=\frac{1}{\pi R}~e^{-2TR^2}.
\ee
Taking $V = V_0 \theta(r_E-r)$, with $T r_E^2\gg 1$, the density perturbation exactly at the surface of Earth is
\be
\frac{\delta \rho}{\rho} = -\frac{V_0}{2}
\int d^3R~G_{\rm MB}(R)= -\frac{V_0}{2T}.
\ee
We have confirmed this result numerically, for a large sphere. This, of course, also follows from an approximate density matrix calculation: on average, the path-integral trajectory spends half the ``time'' inside Earth.

For Fermi-Dirac, 
\be
G_{\rm FD}(R)=\frac{1}{\pi R}~\frac{\sum (-1)^{n+1}\frac{n}{(n^2+4T^2R^2)^2}}{\sum (-1)^{n+1}\frac{1}{n^3}},
\ee
\be
\int d^3R~G_{\rm FD}(R)=\frac{2\ln 2}{3\zeta(3)}\frac{1}{T^2},
\ee
which gives \eqref{eq:drhoFD} for the density exactly at the surface of Earth.
The perturbation theory answer for antineutrinos is the same up to a sign. We have confirmed this result numerically, for a large sphere. 

In our numerical simulations we computed all energy eigenstates inside a sphere of radius 2, with a square potential barrier or well of radius 1, for a particle of mass 1. The eigenstates were computed by separation of angular variables (up to $l=500$). Then the radial coordinate was discretized (up to 1000 zones) and the resulting tridiagonal matrix was numerically diagonalized. In the Maxwell-Boltzmann case, we put $T\gg |V_0|\gg 1$. In the Fermi-Dirac case,  we put $T^2\gg |V_0|\gg 1$. For the well, the bound states were empty, or had occupation numbers equal to 1 (for Fermi-Dirac), or had occupation numbers given by the Maxwell-Boltzmann factor. 

\section{\label{sec:flat} The Flat Earth approximation} 
It often helps to reduce a $3d$ problem to a $1d$ problem; it turns out to be a complication when computing the relic neutrino density around the Earth. The Earth radius $r_E$ is indeed very large as compared with the characteristic momentum of neutrinos $T$ and the critical momentum 
\be
k_* \equiv \sqrt{2V_0}
\ee
at which waves experience $\O(1)$ scattering. This motivates replacing the Earth with a wide $1d$ potential barrier/well of height $V_0$.

Below we will confirm that under such a reduction, the density contrast would still be $\O(V_0)$ but as the result of a delicate cancellation of $\O(\sqrt{V_0})$ contributions. In $1d$, the phase space of modes that are significantly scattered by a wide potential is $\O(\sqrt{V_0})$ and so is the naive expectation for $\delta \rho$. 

\begin{figure}[!]
  \centering
  \includegraphics[scale =.8]{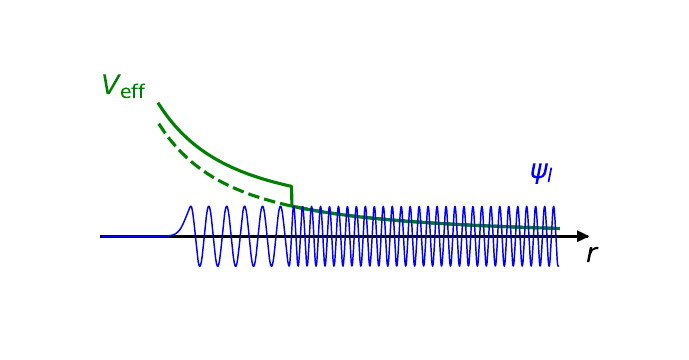} 
  \includegraphics[scale =.8]{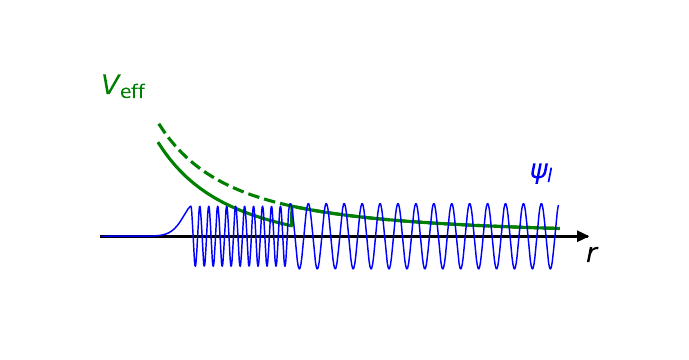} 
  \caption{\small{The effective potential for the radial wavefunctions consists of a step/dip on top of the centrifugal barrier. }}
  \label{fig:Veff}
\end{figure}

Starting from a spherically symmetric $3d$ problem, one arrives at a $1d$ problem by first expanding the exact energy levels in terms of spherical harmonics. The radial wavefunctions satisfy
\be\label{bessel}
(r A_l(r))''+\left[k^2 - 2  V(r) -\frac{l(l+1)}{r^2}\right] r A_l(r) = 0.
\ee
See figure \ref{fig:Veff} for a sketch of the effective potential experienced by $\psi_l = rA_l$. The solutions for $A_l$ are given in terms of different combinations of spherical Bessel functions inside and outside Earth. For a given distribution of momenta at infinity
\be
\rho(r) = \sum_l (2l+1) \int dk k^2 f(k) |A_l(r)|^2,
\ee
and an exact formula for $\delta \rho$ can be written in terms of the Bessel functions \cite{huan}. However, an enormous number of terms, of order $T r_E\sim 10^9$, have to be included for the sum over $l$ to converge. 

One can try to use this large hierarchy to simplify the computation. The first simplification is achieved by switching to $x = r-r_E$, and working at leading order in $x/r_E$ to obtain
\be\label{airy}
\psi_l''(x) + \left[k^2-\frac{l(l+1)}{r_E^2} -2 V(x) + 2\frac{l(l+1)}{r_E^3} x\right]\psi_l(x) = 0,
\ee
where $V(x) = \pm V_0 \theta(-x)$. The solutions are combinations of Airy functions. They are significantly modified with respect to unperturbed solutions when
\be
k_\perp\equiv \sqrt{k^2 - \frac{l(l+1)}{r_E^2}} \sim k_* .
\ee
In the quasi-classical picture, these are (anti)neutrinos that hit the surface of Earth with a small inclination angle $\theta \sim k_*/T$. They might appear to be the main contributors to $\delta \rho$. 

So far we have approximated the centrifugal barrier by a linear potential near the Earth surface. Next, suppose particles with $\kp \sim k_*$ gain a large WKB exponent between $x=0$ and the turning point due to the centrifugal barrier, i.e.
\be \label{wkb}
k_*r_E\gg (l(l+1))^{1/3} \sim (T r_E)^{2/3}.
\ee
Then we can make the further {\em flat Earth approximation}, namely, to replace the effective potential by a wide barrier/well of constant height $\pm V_0$. The centrifugal barrier ensures that all waves are eventually fully reflected, and hence $\psi_l$ can be chosen real, but the exact way this reflection happens does not affect $\rho(x=0)$. 

The condition \eqref{wkb} is not met for the actual values of $r_E,T,k_*$, which give $k_*(r_E/T^2)^{1/3} \sim 1/10$. Geometrically, the inclination angle $\theta\sim k_*/T$ is so small that such particles rapidly exit the Earth \cite{huan}. 

Nevertheless, we proceed by considering fictitious parameters for which the condition holds because it is instructive to see how $\delta \rho = \O(V_0)$ comes about. First, we perform a fully $1d$ calculation
\be\label{drho}
\delta\rho(0) = \int_0^\infty dk f(k) \left(|\psi_{k}(0)|^2-|\psi^{(0)}_k(0)|^2\right),
\ee
where $\psi_k(x)$ and $\psi^{(0)}_k(x)$ are, respectively, the perturbed and unperturbed wavefunctions with energy $k^2/2$. It is easy to embed this calculation in $3d$ by sending $k\to \kp$, $f(k)\to (2l+1) f\left(\sqrt{\kp^2+l(l+1)/r_E^2}\right)$ and summing over $l$. 

In the quasi-classical approximation, the unperturbed wavefunctions rapidly oscillate (as a function of $k$ at fixed $x$) and $|\psi^{(0)}_k(0)|^2$ can be replaced by $1/2$. The perturbed wavefunctions for neutrinos [$V(x)= V_0 \theta(-x)$] can be written as
\be
\psi_{k}(x)  = \left \{\begin{array}{cc}  A(k) \sin\phi_L(k,x),\qquad & x<0,\\
\sin\phi_R(k,x),\qquad & x>0,\end{array}\right.
\ee
for $k>k_*$, and 
\be
\psi_{k}(x)  = \left \{\begin{array}{cc}  A(k) \exp(\phi_L(k,x)),\qquad & x<0,\\
\sin\phi_R(k,x),\qquad & x>0,\end{array}\right.
\ee
for $k<k_*$, where
\be
\d_x \phi_L(k,0) = |\kappa|,\qquad \d_x \phi_R(k,0)= k,\qquad \kappa^2 \equiv k^2 - k_*^2. 
\ee
The wavefunctions and their derivative are continuous at $x=0$. It follows
\be\label{above}
|\psi_k(0)|^2 = \left \{\begin{array}{cc}  \frac{k^2 \sin^2 \phi_L(k,0)}{k^2 \sin^2 \phi_L(k,0)+\kappa^2 \cos^2 \phi_L(k,0)},&\qquad k>k_*,\\[10pt]
\frac{k^2}{k^2 -\kappa^2}, &\qquad k <k_*.\end{array}\right.\ee
Substituting in \eqref{drho}, we have the sum of the following two integrals
\be
I_< = \int_0^{k_*} dk f(k) \frac{2k^2-k_*^2}{2 k_*^2},
\ee
and
\be
I_> = \frac{1}{2}\int_{k_*}^\infty dk f(k) \frac{k^2 \sin^2 \phi_L - \kappa^2 \cos^2\phi_L}{k^2 \sin^2 \phi_L + \kappa^2 \cos^2\phi_L},
\ee
where for brevity we dropped the argument of $\phi_L(k,0)$. For Maxwellian distribution $f(k) = e^{- \frac{k^2}{2T}}$ with $k_*\ll \sqrt{T}$, we find
\be
I_{\rm < } =- \frac{k_*}{6}-\frac{k_*^3}{60  T}+\cdots
\ee
Evaluating $I_>$ is trickier because of the rapid oscillations of the integrand. For a wide barrier $k |\d\phi_L/d k| \gg 1$, so within each oscillation $k$ and $\kappa$ are approximately constant. To use this fact, we change the integration variable to $\phi_L$, split it as
\be
I_> = \frac{1}{2}\sum_n 
\int_{n\pi}^{(n+1)\pi} d\phi_L \frac{d k}{d\phi_L}  f(k) 
\frac{k^2 \sin^2 \phi_L - \kappa^2 \cos^2\phi_L}{k^2 \sin^2 \phi_L + \kappa^2 \cos^2\phi_L},
\ee
and neglect the variation of $k,\kappa, dk/d\phi_l,$ and $f(k)$ inside the integral to find
\be\begin{split}
I_> \approx &\frac{1}{2}\sum_n  \frac{dk}{d\phi_L} f(k)   \pi \frac{(k-\kappa)^2}{k_*^2}\\[10pt]
\approx & \frac{1}{2}\int_{k_*}^\infty dk f(k) \frac{(k-\kappa)^2}{k_*^2}.
\end{split}
\ee
For the same Maxwellian distribution,
\be
I_{> } = -I_< -\frac{\sqrt{\pi}k_*^2}{4\sqrt{2 T}}+\O(k_*/\sqrt{ T})^4,
\ee
while the unperturbed density is $\sqrt{\pi   T/2}$. Hence, to leading order in $V_0$
\be
\frac{\delta\rho(0)}{\rho} = -\frac{V_0}{2T}.
\ee
The result for antineutrinos is the same up to a sign. For them, however, the effective $1d$ potential is a well rather than a barrier, which has $\O(\sqrt{V_0})$ bound states. The occupation numbers of these bound states are seemingly independent of the conditions imposed at infinity, and obviously, the $\O(\sqrt{V_0})$ contributions will not cancel for an arbitrary distribution over these bound states.

Indeed, for an actually $1d$ collisionless gas this distribution depends on the formation history of the potential well. If it is formed adiabatically, then the occupation number of bound states would be the same as the ones in the continuum right above the well and $\delta\rho = \O(V_0)$ \cite{land}. 

Our $1d$ model is instead just an approximation to the $3d$ problem, where the phase space of bound states is $\O(V_0^{3/2})$. When $\sqrt{\kp^2 + l(l+1)/r_E^2} > k_*$, the potential well below the Earth surface is a metastable vacuum (see figure \ref{fig:Veff}). The bound states found in the flat Earth limit are approximations to a subset of the unbound states with large amplitude inside the well and, at equilibrium, their occupation is unambiguously fixed by the distribution at infinity. We will call them quasi-bound states.

To demonstrate the cancellation of $\O(\sqrt{V_0})$ effects, a similar computation as above can be repeated. The unbound wavefunctions are given by
\be
\psi_{k}(x)  = \left \{\begin{array}{cc}  A(k) \sin\phi_L(k,x),\qquad & x<0\\
\sin\phi_R(k,x),\qquad & x>0,\end{array}\right.
\ee
with $k>0$ and 
\be
\d_x \phi_L(k,0) = \sqrt{k^2+k_*^2},\qquad \d_x \phi_R(k,0)= k.
\ee
The quasi-bound states, for a wide potential well can be approximated by a continuum
\be
\psi_{\kappa}(x)  = \left \{\begin{array}{cc} &\sin \Phi_L(\kappa,x),  \qquad x<0\\
&A(\kappa) \exp(-\Phi_R(\kappa,x)),\qquad x>0,\end{array}\right.
\ee
where $0<\kappa<k_*$, and 
\be
\d_x \Phi_L(\kappa,0) = \kappa,\qquad \d_x \Phi_R(\kappa,0)= \sqrt{k_*^2-\kappa^2}.
\ee
Junction conditions at $x=0$ give
\be
|\psi_k(0)|^2 = \left \{\begin{array}{cc}  \frac{k^2 \sin^2 \phi_L(k,0)}{k^2 \sin^2 \phi_L(k,0)+\kappa^2 \cos^2 \phi_L(k,0)},&\qquad \text{unbound},\\[10pt]
\frac{\kappa^2}{k_*^2}, &\qquad \text{quasi-bound.}\end{array}\right.\ee
The variation of $\rho(0)$ differs from \eqref{drho} by the extra contribution of the quasi-bound states with the associated distribution $f_B(\kappa)$
\be
\delta\rho(0) = \int_0^\infty dk f(k) \left(|\psi_{k}(0)|^2-\frac{1}{2}\right) 
+ \int_0^{k_*}d\kappa f_B(\kappa) \frac{\kappa^2}{k_*^2}.
\ee
The first integral is similar to $I_>$. For $f(k) =e^{- \frac{k^2}{2T}}$ with $ k_*\ll \sqrt{T}$,
\be
\delta\rho(0)_{\rm unbound} = -\frac{k_*}{3}+\frac{k_*^2}{8}\sqrt{\frac{\pi}{2 T}}-\frac{k_*^3}{15  T}+\cdots
\ee
whereas assuming the same distribution of the quasi-bound states, namely $f_B(\kappa) =e^{- \frac{\kappa^2-k_*^2}{2T}}$,
\be
\delta\rho(0)_{\rm quasi-bound}= \frac{k_*}{3} +\frac{k_*^3}{15  T}+\cdots
\ee
giving in total $\delta\rho(0)/\rho = V_0/2T$.

Finally, the answer in $3d$ follows from the substitutions described below \eqref{drho} and using the expansion \eqref{eq:FDMB} of Fermi-Dirac distribution in terms of Maxwell-Boltzmann distributions. For every Maxwell-Boltzmann temperature, the distributions of $\kp$ and $l$ factorize, the dependence on $\kp$ being $\exp(-\kp^2/2\alpha T^2)$. Hence, for any $\alpha$, and regardless of $l$, the neutrino/antineutrino density perturbation is
\be
\delta\rho_\alpha = \mp \frac{V_0}{2 \alpha T^2}\rho_\alpha.
\ee
Using $\rho_\alpha\propto \alpha^{3/2}$ and integrating over $\alpha$ gives \eqref{eq:delfd}.

\newpage
\section{Conclusion \label{sec:con}}
The ``Density Theorem" (\S\ref{sec:theo}), the ``thermal calculus'' (\S\ref{sec:Earth}), the perturbation theory (\S\ref{sec:pert}), the brute-force numerical calculation (\S\ref{sec:pert}), the approximate treatment of Bessel functions (\S\ref{sec:flat})---all these point to the conclusion that the density perturbation of relic neutrinos and antineutrinos from the interaction with Earth is small, $\O(G_F)\sim 10^{-8}$. 

Amusingly, perturbation theory and flat Earth approximation have non-overlapping ranges of validity, the former requiring $\ep_1\equiv m_\nu V_0 r_E/T\ll 1$ while the latter $\ep_2\equiv T^2(m_\nu V_0)^{-3/2}/r_E\ll 1$, and neither is satisfied for the actual values of parameters, which give $\ep_1\sim 10^2, \ep_2 \sim 10^3$. Nevertheless, when applicable, they both give the same answer. It is enough to have $T r_E\gg 1$ and $T^2\gg m_\nu V_0$, the conditions upon which the thermal computation relied. 

This is just an example showing how efficient thermal methods are in addressing statistical physics questions. Another, more dramatic example is to consider giving up the assumption of spherical symmetry of Earth, which makes finding the perturbed energy levels cumbersome or utterly impossible while for the thermal method all that matters is the behavior of the potential in a neighborhood of size $1/T$. 

A necessary condition to apply the thermal method to neutrinos, which form an effectively collisionless gas, is that unbound energy levels are occupied as dictated by the distribution at infinity. As we have seen in \S\ref{sec:flat}, if $\ep_2\ll 1$, the potential develops exponentially long-lived quasi-bound states. Hence, the assumption of equilibrium would be invalid if the lifetime of these becomes longer than the age of universe. In reality, however, $\ep_2 \sim 10^3$. 

What led \cite{arva} to a different conclusion? The authors first study the flat Earth limit. They consider a semi-infinite potential, say $V(x<0) = \pm V_0$, and obtain an asymptotic value $\rho(x\to \infty)$ that is different from the unperturbed $\rho$. There is nothing wrong with studying this potential per se, but as a model for Earth it would be more realistic to consider either a finite-size potential barrier/well, or a particle on a half-line (appropriate for the radial wavefunctions for perfectly spherical Earth). Either way the asymptotic density would remain unchanged. Still, they would have obtained an asymmetry of $O(\sqrt{V_0})$ near the surface because they do not include the bound states of antineutrinos in the computation. As discussed in \S\ref{sec:flat}, as a model for the $3d$ problem these are quasi-bound states that have to be included and only then $\delta \rho/\rho = \O(V_0)$. 

Next, the authors consider the spherical Earth limit. In this case, only a geometric optics argument is given based on the fact that neutrinos (but not antineutrinos) with very small inclination angles are reflected by the Earth surface. This is argued to result in an overdensity of neutrinos very close to the surface. Here the error is to forget that any fully reflected neutrino is replacing a neutrino that in the absence of Earth would have directly arrived at the same location but now is being deflected. More generally, by the Liouville theorem there is no effect in the geometric optics approximation. We think this point has been explained in section 3.1 of \cite{huan}.

The expression (15) that \cite{huan} gives for $\delta\rho$ in the $3d$ case is correct. But as acknowledged there, adding $\O(10^9)$ terms (necessary for the sum to converge) is challenging. The paper does not discuss whether the effect is $\O(V_0)$ or $\O(\sqrt{V_0})$, but the numerical answer given there for the sum is too large.

\begin{acknowledgments}
We thank Mina Arvanitaki, Savas Dimopoulos, Ken Van Tilburg and Giovanni Villadoro for useful discussions. 
\end{acknowledgments}

\bibliography{ms}

\end{document}